\documentclass[revtex4-2,twocolumn,showpacs]{openjournal}
\def \be{\begin{equation}}
\def \ee{\end{equation}}
\def \bdm{\begin{eqnarray}}
\def \edm{\end{eqnarray}}

\usepackage{float}
\usepackage{amsmath}
\usepackage{graphicx}

\def \d{\partial}
\newcommand \eqs[1]{Eq. (\ref{#1})}
\newcommand\comment[1]{}
\begin{document}
\title{Subspace Approximation to the Focused Transport Equation. \\II. The Modified Form}
\author{B. Klippenstein and A. Shalchi}
\affiliation{Department of Physics and Astronomy, University of Manitoba, Winnipeg, Manitoba R3T 2N2, Canada}
\email{andreasm4@yahoo.com}
\begin{abstract}
The transport of energetic particles in a spatially varying magnetic field is described by the focused transport equation. In the past two versions
of this equation were investigated. The more commonly used standard form described a pitch-angle isotropization process but does not conserve the norm.
In the current paper we consider the focused transport equation in conservative form also called modified focused transport equation. This equation
conserves the norm but does not describe pitch-angle isotropization. We use the previously developed subspace method to solve the focused transport
equation analytically and numerically. For a pure analytical treatment we employ the two-dimensional subspace approximation. Furthermore, we consider
a higher dimensionality for which one needs to evaluate occurring matrix exponentials numerically. This type of semi-numerical approach is much faster
than traditional solvers and, therefore, it is very useful.
\end{abstract}
%
%
\pacs{47.27.tb, 96.50.Ci, 96.50.Bh}
\maketitle

\section{Introduction}\label{sec:intro}

The theory of energetic particles is an important part of space physics and astrophysics. Solar energetic particles as well as cosmic rays
interact with turbulent magnetic fields leading to a complicated stochastic particle motion (see, e.g., Shalchi (2009a) and Shalchi (2020)
for reviews). Therefore, it is believed that the motion of such particles through interplanetary and interstellar spaces is described
by diffusion coefficients.

In the scenarios described here on assumes that there is a magnetic field consisting of two components. First there is a mean magnetic field
for which one usually assumes $\vec{B}_0 = B_0 \hat{z}$ where $B_0$ is a constant. However, in reality the mean field is not constant. In the
context of space physics the mean field is described by an \textit{Archimedean spiral} (see Parker (1958)). Such a field is complicated and,
therefore, difficult to incorporate into analytical treatments of the transport. An exception are test-particle simulations (see Tautz et al. (2011)).
Thus, one introduces the effect of a non-constant mean field via a so-called focusing term (Roelof (1969)).

Besides the large-scale mean magnetic field there is also a turbulent component. This component causes pitch-angle scattering of the energetic
particles leading to a stochastic particle motion. Therefore, one needs to work with transport equations which are, in all cases, partial
differential equations. At the most fundamental level, this equation would be the Vlasov equation (see Vlasov (1938)). However, this equation
is not very practical and, thus, one can derive an equation for the ensemble average of the particle distribution function (see, e.g., Schlickeiser (2002)
and Zank (2014)). The latter equation is sometimes called the \textit{cosmic ray Fokker-Planck equation} (see Schlickeiser (2002)). Averaging
this transport equation over gyro-phase and pitch-angle leads to a diffusion-convection equation (see, e.g., Parker (1965)).

In the current paper we work with a pitch-angle dependent transport equation only containing pitch-angle scattering and the focusing term.
More complicated and general transport equations can be found in the literature (see, e.g., Skilling (1975), Schlickeiser (2002), and Zank (2014)).
The considered equation is usually called the \textit{focused transport equation} and was investigated analytically in Klippenstein \& Shalchi (2025),
hereafter referred to as Paper I. The standard form of this equation is given by
\be
\frac{\partial f}{\partial t} + v \mu \frac{\partial f}{\partial z}
= \frac{\partial}{\partial \mu} \left[ D_{\mu\mu} \left( \mu \right) \frac{\partial f}{\partial \mu} \right] - \frac{v}{2 L} \big( 1 - \mu^2 \big) \frac{\partial f}{\partial \mu}.
\label{FPeq}
\ee
The latter equation is a partial differential equation with three variables, namely the time $t$, the parallel particle position $z$, and the pitch-angle cosine $\mu$.
Its solution depends on the scattering coefficient $D_{\mu\mu}$ as well as the \textit{focusing length} $L$. The latter parameter corresponds to a characteristic
length scale over which one observes a variation of the mean magnetic field. Eq. (\ref{FPeq}) was explored in several papers such as Earl  (1976), Kunstmann (1979),
Webb (1985), Webb (1987), Ruffolo (1995), Shalchi (2009b), Shalchi \& Danos (2013), Danos et al. (2013), Wang \& Qin (2019), and Wang \& Qin (2020)).

Eq. (\ref{FPeq}) is based on a mean magnetic field of the form
\bdm
B_{0x} \left( x, z \right) & = & \frac{x}{2 L} B_0 e^{-z/L},\nonumber\\
B_{0y} \left( y, z \right) & = & \frac{y}{2 L} B_0 e^{-z/L},\nonumber\\
B_{0z} \left( z \right) & = & B_0 e^{-z/L},
\label{B0zfocused}
\edm
where we have used the constant $B_0$. 

One can easily demonstrate that Eq. (\ref{FPeq}) describes a pitch-angle isotropization process but it does not conserve the norm.  By conservation of magnetic flux,
it is easy to show that the area of a cylindrically symmetric flux tube varies as
\be
A(z) = A(0) e^{z/L}.
\ee
Naturally, the distribution of particles within a flux tube is diluted as the flux tube widens with increasing $z$. It is useful to explicitly take into account the known
shape of the flux tube by considering a different particle distribution function
\be
\tilde{f}(t,z,\mu) = f(t,z,\mu) e^{z/L}.
\label{ftilde}
\ee
This new function is not affected by the geometric dilution of particles. After using Eq. (\ref{ftilde}) in Eq. (\ref{FPeq}) we derive after some straightforward algebra
the focused transport equation in modified form
\be
\frac{\partial \tilde{f}}{\partial t} + v\mu\frac{\partial \tilde{f}}{\partial z}
= \frac{\partial}{\partial \mu}\left[ D_{\mu\mu} \left( \mu \right) \frac{\partial \tilde{f}}{\partial \mu} \right]
- \frac{v}{2L} \frac{\partial}{\partial \mu} \left[(1-\mu^2)\tilde{f}\right].
\label{MFP2}
\ee
The latter equation can be integrated over all positions $z$ and averaged over all $\mu$ to obtain for the norm defined via
\be
\tilde{N} := \frac{1}{2} \int_{-1}^{+1} d \mu \int_{-\infty}^{+\infty} d z \; \tilde{f}(t,z,\mu)
\ee
that $\tilde{N} = const$. This means that Eq. (\ref{MFP2}) indeed conserves the norm. In the current article we use the subspace method to investigate the latter
equation analytically and numerically. This method was described in Lasuik \& Shalchi (2019), Shalchi (2024), and in Paper I.

In Section 2 we combine the subspace method with the modified form of the focused transport equation. Thereafter we approximately solve the Fokker-Planck equation
analytically by employing the two-dimensional subspace method (Section 3). In Section 4 we determine different expectation values. Section 5 discusses the accuracy
and speed of the $N$-dimensional subspace approximation method. In Section 6 we summarize and conclude.

\section{Fourier Transform and Subspace Approximation}\label{modifiedequation}

To solve Eq. (\ref{MFP2}) via the $N$-dimensional subspace method, we first rewrite it by using a Fourier representation of the form
\be
f \big( z, \mu, t \big) = \int_{-\infty}^{+\infty} d k_{\parallel} \;  F_{k_\parallel} \big(\mu, t \big) e^{i k_{\parallel} z}.
\label{FourierTransform}
\ee
Here and during the reminder of this paper we omit the tilde notation meaning that the functions $f$ and $F_{k_\parallel}$ are the solutions of the modified
equation. Later we shall also use the inverse transform which is given by
\be
 F_{k_\parallel} \big(\mu, t \big) = \frac{1}{2 \pi} \int_{-\infty}^{+\infty} d z \; f \big( z, \mu, t \big) e^{- i k_{\parallel} z}.
\label{invFourier}
\ee
Using this representation in Eq. (\ref{MFP2}) yields
\bdm
& & \frac{\partial F_{k_\parallel}}{\partial t} + i v \mu k_{\parallel} F_{k_\parallel} \nonumber\\
& = & \frac{\partial}{\partial \mu} \left[ D_{\mu\mu} \frac{\partial F_{k_\parallel}}{\partial \mu} \right]
- \frac{v}{2 L} \frac{\partial}{\partial \mu} \left[ \big( 1 - \mu^2 \big) F_{k_\parallel} \right].
\label{FourierFPeq}
\edm
As in Paper I we expand the function $F_{k_\parallel}$ in Legendre polynomials $P_n (\mu)$ so that
\be
F_{k_\parallel} (\mu,t) = \sum_{n=0}^{\infty} C_n(t) P_n (\mu).
\label{expandF}
\ee
Expanding the solution of pitch-angle dependent transport equations is well-known in the literature (see, e.g. Earl (1974) and Zank et al. (2000b)).
Furthermore, we need to specify the form of the scattering coefficient $D_{\mu\mu}$. As in Paper I we employ the so-called \textit{isotropic form}
\be
D_{\mu\mu} = D (1 - \mu^2)
\ee
which was derived systematically in Shalchi et al. (2009) based on the \textit{second-order quasi-linear theory} developed in Shalchi (2005). This form
should be valid for strong ($\delta B \gg B_0$) and intermediate strong ($\delta B \approx B_0$) turbulence as found in interplanetary and interstellar
spaces. This isotropic form is often used in analytical investigations of pitch-angle scattering equations (see, e.g., Malkov \& Sagdeev (2015),
Malkov (2015), and Malkov (2017)).

With this form and the expansion given above, Eq. (\ref{FourierFPeq}) turns into
\bdm
& & \sum_{n}\dot{C}_n P_n+iv\mu k_\parallel\sum_{n} C_n P_n \nonumber\\
& = & D \sum_{n} C_n \frac{\partial}{\partial \mu} \left[ \left( 1 - \mu^2 \right) \frac{\partial P_n}{\partial \mu} \right]\nonumber\\
& - & \frac{v}{2 L}\sum_{n} C_n\frac\d{\d\mu}\left[(1-\mu^2)P_n\right].
\label{MFP3}
\edm
To rewrite the last term, we use the relation (see Abramowitz \& Stegun (1968))
\be
\frac\d{\d\mu} \left[ (1-\mu^2) P_n \right] = (n-1)\mu  P_n-(n+1)P_{n+1}.
\ee
Combining this with 
\be
\mu P_n = \bigg( \frac{n+1}{2n+1} \bigg) P_{n+1} + \bigg( \frac{n}{2n+1} \bigg) P_{n-1},
\label{recurrence}
\ee
yields
\bdm
\frac\d{\d \mu}\left[(1-\mu^2)P_n\right] & = & (n+1)\left[\frac{n-1}{2n+1}-1\right]P_{n+1}\nonumber\\
& + & \frac{(n-1)n}{2n+1}P_{n-1}.
\edm
In the last term of Eq. (\ref{MFP3}) we use
\be
\frac{\partial}{\partial \mu} \left[ \left( 1 - \mu^2 \right) \frac{\partial P_n}{\partial \mu} \right] = - n \left( n + 1 \right) P_n.
\label{Legendresdiffeqnwithn}
\ee
Using the relations listed above, allows us to write Eq. (\ref{MFP3}) as
\bdm
& & \sum_{n}\dot{C}_n P_n + i v k_\parallel \sum_{n} C_n \bigg[ \frac{n+1}{2n+1} P_{n+1} + \frac{n}{2n+1} P_{n-1} \bigg] \nonumber\\
& = & - D\sum_{n} C_n n (n+1) P_n \nonumber\\
& - & \frac{v}{2 L} \sum_{n} C_n \left[ (n+1)\left(\frac{n-1}{2n+1}-1\right)P_{n+1} \right.\nonumber\\
& + & \left. \frac{(n-1)n}{2n+1}P_{n-1} \right].
\edm
To continue we multiply the latter equation by $P_m$ and integrate over all $\mu$. Thereafter we use the orthogonality relation
\be
\int_{-1}^{+1} d \mu \; P_n P_m = \frac{2}{2 m + 1} \delta_{nm},
\label{Pnortho}
\ee
and obtain
\bdm
\dot{C}_m & = & - ivk_\parallel\left[\frac m{2m-1} C_{m-1}+\frac{m+1}{2m+3} C_{m+1} \right] \nonumber\\
& - & \frac v{2L}\left[-\frac{m(m+1)}{2m-1} C_{m-1}+\frac{m(m+1)}{2m+3} C_{m+1}\right] \nonumber\\
& - & D m (m+1) C_m.
\label{MCE}
\edm
This corresponds to and infinite number of coupled differential equations for the function of time $C_m$. It can be written as the matrix equation
\be
\dot{C}(t)  = \boldsymbol{M} C(t)
\label{thematrixequation}
\ee
where we have used the column vector 
\be
C(t) = \left(\begin{array}{c}
C_0(t)      \\
C_1(t) 	 \\
\vdots 	 \\
\end{array}\right).
\ee
and the elements of the matrix $\boldsymbol{M}$ are given as the coefficients of $C_n$ in Eq. (\ref{MCE}). The formal solution of this equation
can be written as
\be
C(t) = e^{\boldsymbol{M} t} C(0)
\label{tildecn_exact}
\ee
where we have used the matrix exponential. Note, this formulation is exactly as in Paper I but now we have a different matrix $\boldsymbol{M}$.
Assuming sharp initial conditions yields
\be
C_m \big( t = 0 \big) = \frac{2 m + 1}{2 \pi} P_m \big( \mu_0 \big)
\label{InitialCoeffRelation}
\ee
as demonstrated in Paper I. While Eq. (\ref{tildecn_exact}) is indeed the solution to our problem, the matrix exponential cannot be evaluated
exactly. One way of approaching this problem is to employ the \textit{$N$-dimensional subspace approximation}. In this case we cut off the expansion
given by Eq. (\ref{expandF}) by keeping $N$ terms. After doing this, the matrix $\boldsymbol{M}$ is approximated by an $N \times N$ matrix.

For small values of $N$ such as $N=2$ we can then find an analytical solution to our problem (see below). However, in order to find very accurate solutions,
large values of $N$ are required. In this case a pure analytical solution is no longer possible, but we can evaluate the matrix exponential by using software
such as MATLAB.

\section{Two-Dimensional Subspace Approximation}

We start our investigations by considering the case $N=2$ corresponding to the two-dimensional subspace approximation. After truncating
the expansion given by Eq. (\ref{expandF}) it becomes
\be
F_{k_\parallel} (\mu,t) = C_0 (t) + C_1 (t) \mu
\label{cutofat2}
\ee
By setting $m=0$ in \eqs{MCE}, we can see that
\be
\dot {C}_0 = -i\frac13vk_\parallel C_1.
\label{MC0}
\ee
Similarly, if we set $m=1$ in Eq. (\ref{MCE}), we derive
\be
\dot {C}_1= -ivk_\parallel C_0 -2D C_1+\frac vL C_0
\label{MC1}
\ee
where we have set $C_2 = 0$. Combining \eqs{MC0} with \eqs{MC1} results in a linear system that we can write as the matrix equation
\bdm
\left(
\begin{array}{c}
\dot{C}_0      \\[0.5cm]
\dot{C}_1      
\end{array}
\right) = \left(
\begin{array}{ccc}
0   							\quad & - \frac{1}{3} i v k_{\parallel} 		\\[0.5cm]
\frac vL - i v k_{\parallel}	\quad & - 2 D										
\end{array}
\right) \left(
\begin{array}{c}
C_0      \\[0.5cm]
C_1 	
\end{array}
\right)
\label{MS}
\edm
which is a special case of Eq. (\ref{thematrixequation}). To solve this, we require the eigenvalues of \eqs{MS}. It is straightforward to show they solve
\be
\omega^2+2D \omega+\frac13 v^2k_\parallel^2+i\frac13\frac{v^2k_\parallel}L=0
\ee
and are given by
\be
\omega_\pm =-D\pm \sqrt{D^2-\frac13v^2k_\parallel^2-i\frac{v^2k_\parallel}{3L}}.
\label{coefficientsomegapm}
\ee
Now we suppose that $C_0$ has the form
\be
C_0 (t)= b_+ e^{\omega_+t} + b_- e^{\omega_-t}.
\label{MC0t}
\ee
Placing \eqs{MC0t} into \eqs{MC0} yields
\be
C_1 (t)=i\frac3{vk_\parallel} \left[ \omega_+ b_+e^{\omega_+t} + \omega_- b_-e^{\omega_-t}\right].
\label{MC1t}
\ee
To determine $b_\pm$, we use the initial conditions provided by Eq. (\ref{InitialCoeffRelation}) and find
\bdm
b_+ + b_-& = & \frac1{2\pi}\nonumber\\
i\frac3{vk_\parallel}\left(\omega_+ b_+ + \omega_- b_-\right) &=& \frac{3\mu_0}{2\pi}.
\edm
After some straightforward algebra, one can show that
\bdm
b_+ & = & -\frac1{2\pi}\frac{ \omega_-+iv\mu_0k_\parallel}{\omega_+ - \omega_-}\nonumber\\
b_- & = & \frac1{2\pi}\frac{\omega_+ + iv\mu_0k_\parallel}{\omega_+ - \omega_-}.
\label{coefficientsbpm}
\edm
The coefficients given by Eqs. (\ref{MC0t}) and (\ref{MC1t}) can be used in Eq. (\ref{cutofat2}) to write
\be
F_{k_\parallel}(\mu,t) = \left(1+i\frac{3 \omega_+\mu}{vk_\parallel}\right) b_+e^{\omega_+t} +\left(1+i\frac{3 \omega_-\mu}{vk_\parallel}\right) b_-e^{\omega_-t}
\ee
where the coefficients $b_{\pm}$ are given by Eq. (\ref{coefficientsbpm}) and the coefficient $\omega_{\pm}$ are provided by Eq. (\ref{coefficientsomegapm}).

\section{Expectation Values}

In the following we determine different expectation values. The ensemble average is defined via
\bdm
\big< A \big> & = & \frac{1}{4} \int_{-1}^{+1} d \mu \; \int_{-1}^{+1} d \mu_0 \nonumber\\
& \times & \int_{-\infty}^{+\infty} d z \; A \left( z, \mu, t \right) f \big( z, \mu, t \big).
\label{defaverageoperator}
\edm
In some cases a pitch-angle dependent solution is desired. In such cases we can drop one of the integrals in Eq. (\ref{defaverageoperator}).
In the next few sections we consider some examples for $A$. Some of those expectation values are needed to develop analytical
theories for perpendicular transport (see, e.g., Matthaeus et al. (2003), Shalchi (2010), Shalchi (2020), and Shalchi (2021)).

\subsection{The Characteristic Function}

As a first application, we compute the characteristic function defined via
\bdm
\left< e^{- i k_{\parallel} z} \right> & = & \frac{1}{4} \int_{-1}^{+1} d \mu \int_{-1}^{+1} d \mu_0 \nonumber\\
& \times & \int_{-\infty}^{+\infty} d z \; f \big( z, \mu, t \big) e^{- i k_{\parallel} z}.
\label{eq:characteristic}
\edm
Therein we can use Eq. (\ref{invFourier}) to obtain
\be
\left< e^{- i k_{\parallel} z} \right> = \frac{\pi}{2} \int_{-1}^{+1} d \mu \int_{-1}^{+1} d \mu_0 \; F_{k_\parallel} \big(\mu, t \big)
\ee
meaning that the desired characteristic function is the $\mu_0$- and $\mu$-averaged function $F_{k_\parallel}$. If we average Eq. (\ref{expandF})
over all $\mu$, we find
\be
\frac{1}{2} \int_{-1}^{+1} d \mu \;  F_{k_\parallel} \big(\mu, t \big) = C_0 \big( t \big)
\label{eq:Fint}
\ee
where we have used orthogonality relation of the Legendre polynomials provided by Eq. (\ref{Pnortho}). With Eq. (\ref{MC0t}) for the function $C_0 (t)$ we
find for the characteristic function
\be
\left< e^{- i k_{\parallel} z} \right> = \pi \int_{-1}^{+1} d \mu_0 \; \left[ b_+ e^{\omega_+ t} + b_- e^{\omega_- t} \right].
\ee
Using therein Eq. (\ref{coefficientsbpm}) and solving the $\mu_0$-integral yields
\be
\left< e^{- i k_{\parallel} z} \right> = \frac{\omega_+}{\omega_+ - \omega_-}e^{\omega_-t}-\frac{\omega_-}{\omega_+ - \omega_-}e^{\omega_+ t}.
\label{eq:modifiedchar}
\ee

The characteristic function investigated so far is related to the $\mu_0$- and $\mu$-averaged distribution function via
\be
M \left( z, t \right) = \frac{1}{2 \pi} \int_{-\infty}^{+\infty} d k_{\parallel} \; \langle e^{\pm i k_{\parallel} z} \rangle e^{i k_{\parallel} z}.
\label{FourierChar}
\ee
Therein we can use Eq. (\ref{eq:modifiedchar}) to find $M \left( z, t \right)$.

In the following paragraphs, we aim to simplify this result for the extremal values of the wave number.

\subsubsection{Small Wave Numbers}

First we consider small wave numbers and assume
\be
\left|\frac13v^2k_\parallel^2+i\frac{v^2k_\parallel}{3L}\right| \ll D^2.
\label{eq:modifiend_small}
\ee
If we write
\be
\omega_\pm = -D \pm \sqrt{\dots},
\ee
we have
\be
\sqrt{\dots} \approx D - i \frac{\kappa_\parallel}{L}k_\parallel - \bar\kappa_\parallel k_\parallel^2
\ee
where we have used the parallel spatial diffusion coefficient without focusing $\kappa_\parallel$ given by (see Shalchi (2006))
\be
\kappa_{\parallel} = \frac{v^2}{6 D}
\label{eq:kparallel}
\ee
and $\bar\kappa_\parallel$ given in 
\be
\bar{\kappa}_{\parallel} = \kappa_{\parallel} \left( 1 - \frac{1}{3} \frac{\lambda_{\parallel}^2}{L^2} \right)
\label{defkappabar}
\ee
where we have used the parallel mean free path without focusing
\be
\lambda_{\parallel} = \frac{v}{2 D}.
\label{eq:lambdamean}
\ee
This result is very similar compared to the formulas derived in Wang \& Qin (2020) as well as Shalchi \& Klippenstein (2025) and identical compared to the result obtained
in Paper I.

With the approximation described above we can write
\be
\omega_+=-i\frac{\kappa_\parallel}{L}k_\parallel-\bar\kappa_\parallel k_\parallel^2\label{eq:op_small}
\ee
and
\be
\omega_-=-\frac{v}{\lambda_\parallel}+i\frac{\kappa_\parallel} L k_\parallel.
\ee
Therefore, we can write Eq. (\ref{eq:modifiedchar}) as
\bdm
\left< e^{- i k_{\parallel} z} \right>
& = & \frac{-i\frac{\kappa_\parallel}{L}k_\parallel-\bar\kappa_\parallel k_\parallel^2}{2\sqrt{\dots}}e^{-vt/\lambda_\parallel+i\kappa_\parallel k_\parallel t/L}\nonumber\\
& + & \frac{\frac{v}{\lambda_\parallel}-i\frac{\kappa_\parallel}Lk_\parallel}{2\sqrt{\dots}}e^{-i\kappa_\parallel k_\parallel t/L-\bar\kappa_\parallel k_\parallel^2t}.
\edm
We can further simplify by noticing that the first term decays much quicker than the second term.  Thus, we obtain
\be
\left< e^{- i k_{\parallel} z} \right>=e^{-i\kappa_\parallel k_\parallel t/L-\bar\kappa_\parallel k_\parallel^2t}\label{eq:char_small_wavenumber}
\ee
which is identical to the result for the unmodified equation (see again Paper I).

\subsubsection{Large Wave Numbers}

For large wave numbers, on the other hand, we assume
\be
v^2k_\parallel^2 \gg \left| -3D^2 +i\frac{v^2k_\parallel}L\right|.
\ee
This implies that we can approximate
\bdm
\sqrt{\dots}
& \approx & i \frac{v|k_\parallel|}{\sqrt3}-\frac{v\text{sgn}(k_\parallel)}{2\sqrt3 L}-i\frac{\sqrt3 D^2}{2v|k_\parallel|} \nonumber\\
& \approx & - \frac{v\text{sgn}(k_\parallel)}{2\sqrt3 L} + i \frac{v|k_\parallel|}{\sqrt3}.
\edm
We obtain
\be
\omega_+ =-D -\frac{v\text{sgn}(k_\parallel)}{2\sqrt3 L}+i\frac{v|k_\parallel|}{\sqrt3},
\ee
\be
\omega_- =-D +\frac{v\text{sgn}(k_\parallel)}{2\sqrt3 L}-i\frac{v|k_\parallel|}{\sqrt3},
\ee
and
\be
\omega_+ - \omega_-=2\sqrt{\dots} \approx -\frac{v\text{sgn}(k_\parallel)}{\sqrt3 L}+i\frac{2v|k_\parallel|}{\sqrt3}.
\ee
We can then simplify Eq.  (\ref{eq:modifiedchar}) by
\bdm
\left< e^{- i k_{\parallel} z} \right>
& = & \frac1{-\frac1{2k_\parallel L}+i} \Bigg[ \left(-\frac{\sqrt3}2\frac{D}{v|k_\parallel|}-\frac1{4k_\parallel L}+i\frac12\right) \nonumber\\
& \times & e^{-\left( \frac{v\text{sgn}(k_\parallel)}{2\sqrt3 L} + i\frac{v|k_\parallel|}{\sqrt3}\right)t} \nonumber\\
& + & \left(\frac{\sqrt3}2\frac{D}{v|k_\parallel|}-\frac1{4k_\parallel L}+i\frac12\right) \nonumber\\
& \times & e^{\left(\frac{v\text{sgn}(k_\parallel)}{2\sqrt3 L}+i\frac{v|k_\parallel|}{\sqrt3}\right)t} \Bigg] e^{-Dt}.
\edm
If we again assume that $L,|k_\parallel|\to\infty$, we obtain the same result as in Paper I, namely
\be
\left< e^{- i k_{\parallel} z} \right> = \cos\left(\frac{v|k_\parallel|}{\sqrt3}t\right)e^{-Dt}
\ee
corresponding to a damped unperturbed motion (see, e.g., Lasuik \& Shalchi (2019)).

\subsubsection{Early and Late Times}

Thanks to the above simplifications, we can obtain a pleasant result for $M(z,t)$ given by Eq. (\ref{FourierChar}) for early and late times.  

First, for the initial value, we place Eq.  (\ref{eq:modifiedchar}) at $t=0$ into Eq. (\ref{FourierChar}) to obtain
\be
M(z,t=0) = \frac1{2\pi}\int_{-\infty}^{+\infty} dk_\parallel e^{ik_\parallel z}.
\label{eq:M_initial}
\ee
This is a well-known integral that can be found in \cite{zwillinger12} and is given as
\be
\int_{-\infty}^{+\infty} d k_\parallel \; e^{i k_{\parallel} z} = 2 \pi \delta (z ).
\label{zwillinger}
\ee
Placing this into Eq. (\ref{eq:M_initial}) yields the expected result of
\be
M(z,t=0) = \delta(z)
\ee
as we have assumed sharp initial conditions at $z=0$.

As for the late time limit,  we have that unless the wavenumber is small,  $e^{\omega_\pm t}$ is effectively zero.  This allows us to approximate the characteristic function in Eq.  (\ref{FourierChar}) with Eq.  (\ref{eq:char_small_wavenumber}).  We therefore have
\bdm
M(z,t\to\infty) &=& \frac1{2\pi} \int_{-\infty}^{+\infty} dk_\parallel e^{i k_\parallel (z-\kappa_\parallel t/L)-\bar\kappa_\parallel k_\parallel^2t}\nonumber\\
&=&\frac1{\sqrt{4\pi \bar\kappa_\parallel t}}e^{-(z-\kappa_\parallel t/L)^2/(4\bar\kappa_\parallel t)}.
\edm
Since this is given in terms of a Gaussian distribution,  we readily get the values of the mean position and the mean square displacement:
\bdm
\left<z\right>(t\to\infty) &=& \frac{\kappa_\parallel }Lt\label{eq:exp_z_char}\\
\left[\left<z^2\right> - \left<z\right>^2\right](t\to\infty)&=& 2\bar\kappa_\parallel t.\label{eq:msd_char}
\edm
By combining these two results, we can also see that
\be
\left< z^2\right>(t\to\infty) = \frac{\kappa_\parallel^2}{L^2}t^2.\label{eq:exp_z2_char}
\ee
It is interesting to note that these results are identical to those derived in Paper I, except the mean position has the opposite sign. 

\subsection{The Velocity Correlation Function}

According to the TGK (Taylor-Green-Kubo) formulation (see Taylor (1922), Green (1951), and Kubo (1957)), a parallel diffusion coefficient can be computed
by using the formula
\be
\kappa_{\parallel}^{TGK} = \int_{0}^{\infty} d t \; \left< V_z (t) V_z (0) \right>.
\label{tgk}
\ee
Alternatively, a diffusion coefficient can be defined via the moments and so that
\be
\bar{\kappa}_{\parallel} \equiv \kappa^{MSD}_{\parallel} = \frac{1}{2} \frac{d}{d t} \left[ \big< z^2 \big> - \big< z \big>^2 \right].
\label{zMSD}
\ee
It is often thought that both definitions yield the same diffusion coefficient but this is not true in a spatially varying mean magnetic field
(see Danos et al. (2013) as well as Shalchi \& Klippenstein (2025)).

In the following we determine the velocity correlation function
\be
V_{zz} (t) = \left< V_z (t) V_z (0) \right>
\ee
controlling the diffusion coefficient $\kappa_{\parallel}^{TGK}$. The velocity correlation function is obtained via
\be
V_{zz}(t) = \frac{v^2}3\left[\frac{\omega_+}{\omega_+ - \omega_-}e^{\omega_+ t}-\frac{\omega_-}{\omega_+ - \omega_-}e^{\omega_-t}\right]_{k_\parallel=0}
\ee
as demonstrated in Paper I. Again, we have for $k_\parallel=0$
\be
\omega_+ = 0 \quad\textnormal{and}\quad \omega_- = - 2 D.
\ee
Therefore, we obtain
\be
V_{zz}(t) = \frac{v^2}{3} e^{- 2 D t}
\ee
as found in Paper I. This corresponds not only to the previously found result but also to the formula derived in Shalchi (2006) for the case
without focusing. Thus, the parameter $\kappa_{\parallel}^{TGK}$ is equal to the case without focusing. This is a consequence of the 
employed subspace approximation (see Shalchi \& Klippenstein (2025) for exact results for the different parallel diffusion coefficients
in a spatially varying magnetic field).

\subsection{The Expectation Value $\left< \mu \right>$}

As in Paper I we determine the moment $\left< \mu \right>$. The latter parameter is calculated by employing
\be
\left< \mu \right> = \frac{1}{2} \int_{-1}^{+1} d \mu \int_{-\infty}^{+\infty} d z \; \mu f \left( z, \mu, t \right).
\ee
Using therein Eq. (\ref{FourierTransform}) yields
\be
\left< \mu \right> = \pi \int_{-1}^{+1} d \mu \; \mu F_0 \left( \mu, t \right).
\ee
Employing therein Eq. (\ref{Pnortho}) results in
\be
\left< \mu \right> = \frac{2}{3} \pi C_1 (k_\parallel=0, t).
\ee
Thanks to the factor $k_\parallel$ in the denominator of Eq. (\ref{MC1t}), evaluating $\left<\mu\right>$ is more difficult than simply setting $k_\parallel=0$.
Instead, one must proceed using L'Hopital's rule. With this in mind, a straightforward calculation shows
\be
\left< \mu \right> = \frac v{6DL}+ \left (\mu_0 - \frac{v}{6DL}\right) e^{-2Dt} .
\ee
As demonstrated, we have that $\left< \mu \right>$ does not in fact decay to zero. For $t \rightarrow \infty$ we find
\be
\left< \mu \right> = \frac{v}{6 D L} = \frac{1}{3} \frac{\lambda_{\parallel}}{L}
\ee
where we have used Eq. (\ref{eq:lambdamean}). The results for the moment $\left< \mu \right>$ is very different compared to the formula obtained in Paper I.
The formula derived here gives us for the so-called \textit{coherent speed}
\be
v_c = v \left< \mu \right> = \frac{\kappa_{\parallel}}{L}
\ee
as found, for instance, in Shalchi \& Danos (2013).

\subsection{The Moments $\left< z^n \right>$}

The final set of expectation values that we consider are the moments $\left< z^n \right>$.  These are obtained via
\bdm
\left<z^n\right> & = & \frac12 \int_{-1}^{+1} d \mu  \int_{-\infty}^{+\infty} dz \; z^n f (z,\mu,t)\nonumber\\
& = & \frac12 \int_{-1}^{+1} d\mu \; \int_{-\infty}^{+\infty} d k_\parallel \; F_{k_\parallel} (\mu,t) \int_{-\infty}^{+\infty} dz \; z^n e^{ik_\parallel z}\nonumber\\
\label{zn}
\edm
where we have used Eq.  (\ref{FourierTransform}). Using
\be
z^n e^{i k z} = (-i)^n \frac{\partial^n}{\partial k_\parallel^n} e^{i k_\parallel z}
\ee
in Eq. (\ref{zn}) yields
\bdm
\left< z^n \right> 
& = & (-i)^n \frac12 \int_{-1}^{+1} d\mu \int_{-\infty}^{+\infty} d k_\parallel \; F_{k_\parallel} (\mu,t)\nonumber\\
& \times & \int_{-\infty}^{+\infty} dz \; \frac{\partial^n}{\partial k_\parallel^n} e^{i k_\parallel z}.
\label{zn2}
\edm
Integrating by parts $n$-times gives us
\bdm
\left< z^n \right> & = & i^n \frac{1}{2} \int_{-1}^{+1} d \mu \int_{-\infty}^{+\infty} d k_\parallel \; \frac{\partial^n}{\partial k_\parallel^n} F_{k_\parallel} (\mu,t) \nonumber\\
& \times & \int_{-\infty}^{+\infty} d z \; e^{i k_\parallel z}.
\edm
The $z$-integral is provided by Eq.  (\ref{zwillinger}) and results in
\be
\left< z^n \right> = i^n \pi \int_{-1}^{+1} d \mu \int_{-\infty}^\infty d k_\parallel \; \delta (k_\parallel) \frac{\partial^n}{\partial k_\parallel^n} F_{k_\parallel} (\mu,t).
\ee
The integral over the Dirac delta can be evaluated and we find
\be
\left< z^n \right> = i^n \pi \left[ \frac{\partial^n}{\partial k_\parallel^n} \int_{-1}^{+1} d\mu \; F_{k_\parallel} (\mu,t) \right]_{k_\parallel=0}.
\label{zn_simplified}
\ee
To evaluate this further we use Eq. (\ref{expandF}) and employ the orthogonality relation provided by Eq. (\ref{Pnortho}) to obtain
\be
\left< z^n \right> = i^n 2 \pi \left[ \frac{\partial^n}{\partial k_\parallel^n} C_0 \left( k_\parallel, t \right) \right]_{k_\parallel=0}.
\label{eq:exzn}
\ee
Attempting to solve this exactly, even for $n=1$ and $2$, only results in an expression which is too complicated to be of much interest here.
Hence, we elect to only consider the limiting values in late times. Once more, the calculations here are similar compared to Paper I, and thus
we refer to that for further details of the calculations.  

We begin with noticing that by Eq.  (\ref{coefficientsomegapm}),  $\omega_+(k_\parallel=0)=0$ which implies that for late times and small wave
numbers, Eq. (\ref{MC0t}) reduces down to
\be
C_0(t) = b_+e^{\omega_+t}.
\label{eq:c0_late_small}
\ee
For further simplifications, we must specify the value of $n$. In the rest of this section, we deal with the cases of $n=1$ and $n=2$.

For $n=1$, Eqs. (\ref{eq:exzn}) and (\ref{eq:c0_late_small}) in late times give
\be
\left<z\right>(t\to\infty) = i2\pi \left[b_+t\frac\d{\d k_\parallel} \omega_+\right]_{k_\parallel=0}.
\ee
Evaluating the derivatives and setting $k_\parallel=0$ results in
\be
\left<z\right>(t\to\infty) = \frac{\kappa_\parallel}Lt.
\label{eq:exp_z}
\ee

Next, for $n=2$, we take the second derivative of Eq. (\ref{eq:c0_late_small}) and keep only the highest order terms in time to derive
\be
\left<z^2\right>(t\to\infty)=-2\pi \left[b_+\left(\frac{d}{dk_\parallel}\omega_+\right)^2t^2\right]_{k_\parallel=0}.
\ee
Computing the derivative and setting $k_\parallel=0$ gives us
\be
\left<z^2\right>(t\to\infty)=\frac{\kappa_\parallel^2}{L^2}t^2.
\label{eq:exp_z2}
\ee
Eq. (\ref{eq:exp_z}) agrees with Eq. (\ref{eq:exp_z_char}), and similarly for Eqs. (\ref{eq:exp_z2}) and (\ref{eq:exp_z2_char}).

\section{$N$-Dimensional Subspace Approximation}

Like in Paper I we now develop the $N$-dimensional subspace method corresponding to a semi-analytial/semi-numerical approach. This is a very efficient method
do solve pitch-angle dependent transport equations (see Shalchi (2024) and Paper I). It is in particular very fast and accurate if one wants to obtain
numerically the moments $\langle z \rangle$ and $\langle z^2 \rangle$ as well as the diffusion coefficient $\kappa^{MSD}_{\parallel}$ (see Shalchi \& Klippenstein (2025)).

As described in Sect. \ref{modifiedequation}, within the $N$-dimensional subspace approximate we cut off the expansion provided by Eq. (\ref{expandF}) so that we only keep
the first $N$ terms therein. Therefore, we need to solve the matrix exponential given by Eq. (\ref{tildecn_exact}) for the case that $\boldsymbol{M}$ is an $N \times N$ matrix.
This step has to be done numerically. As in Paper I we go up to $N=10$ since it provides an accurate solution within a short computational time.

For comparison we also solve the considered focused transport equation by employing an \textit{implicit Euler method} corresponding to a pure numerical approach
(see Shalchi (2024) for details). This allows us to test the subspace method for different values of $N$.

\begin{figure}[H]
\centering
\includegraphics[scale=0.5]{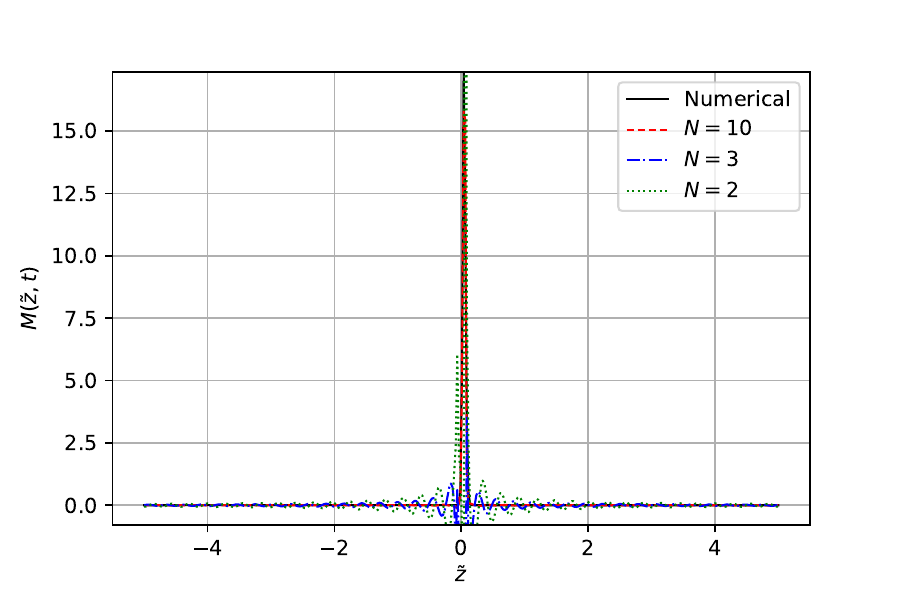}
\caption{This plot compares the $\mu$-integrated solution of the subspace method for various dimensions and the numerical solution at the time $t=1/10$ for an
initial value $\mu_0=1/2$. We have set the focusing parameter $\xi = 1/5$.}
\label{modified_mu_integrated_5_01}
\end{figure}

\begin{figure}[H]
\centering
\includegraphics[scale=0.5]{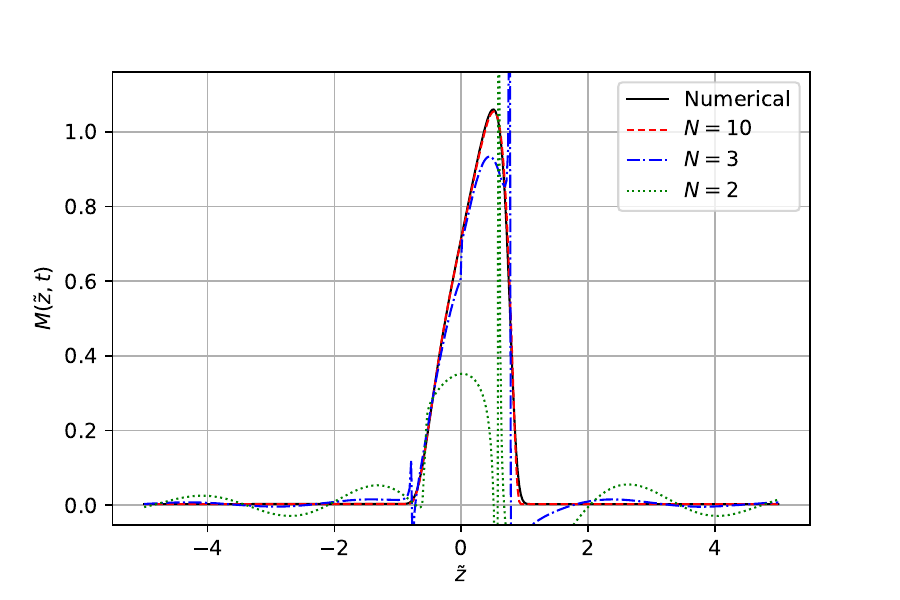}
\caption{Caption is as in Figure \ref{modified_mu_integrated_5_01} except that we have used $t =1$.}
\label{modified_mu_integrated_5_1}
\end{figure}

\begin{figure}[H]
\centering
\includegraphics[scale=0.5]{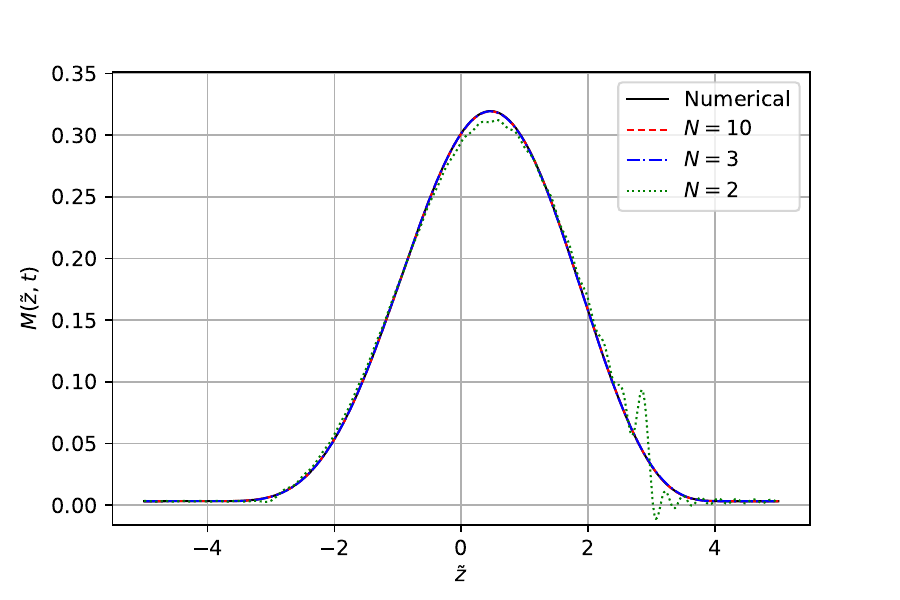}
\caption{Caption is as in Figure \ref{modified_mu_integrated_5_01} except that we have used $t =5$.}
\label{modified_mu_integrated_5_5}
\end{figure}

\begin{figure}[H]
\centering
\includegraphics[scale=0.5]{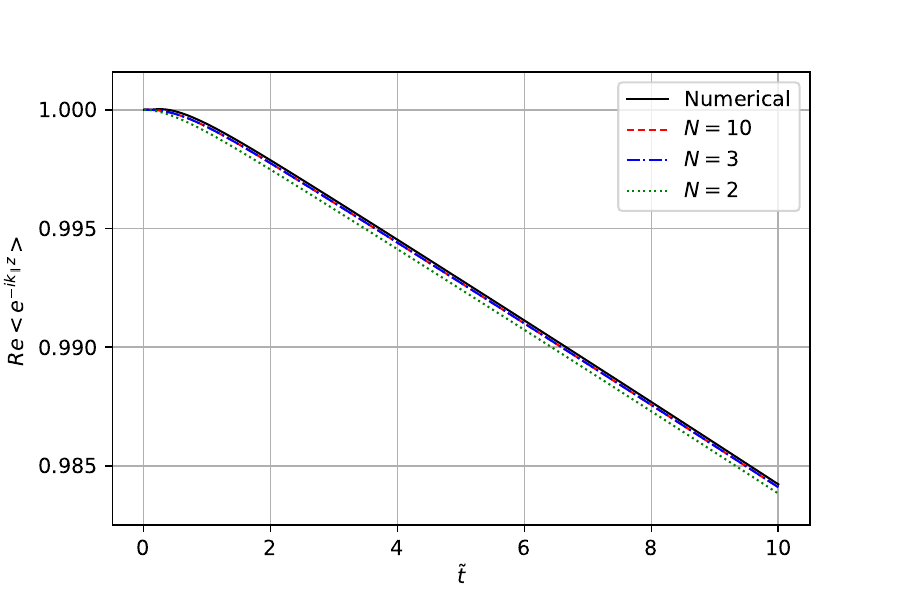}
\caption{Comparison of the $\mu$-averaged characteristic function between the subspace method for various dimensions and the numerical solution. We use an
initial value $\mu_0=0$ and $\xi =1/5$ for the focusing parameter. Here, we keep the dimensionless wave number $\tilde{k} = v k / D = 1/10$ constant and
varied time.}
\label{modified_char_5_k_01}
\end{figure}

\begin{figure}[H]
\centering
\includegraphics[scale=0.5]{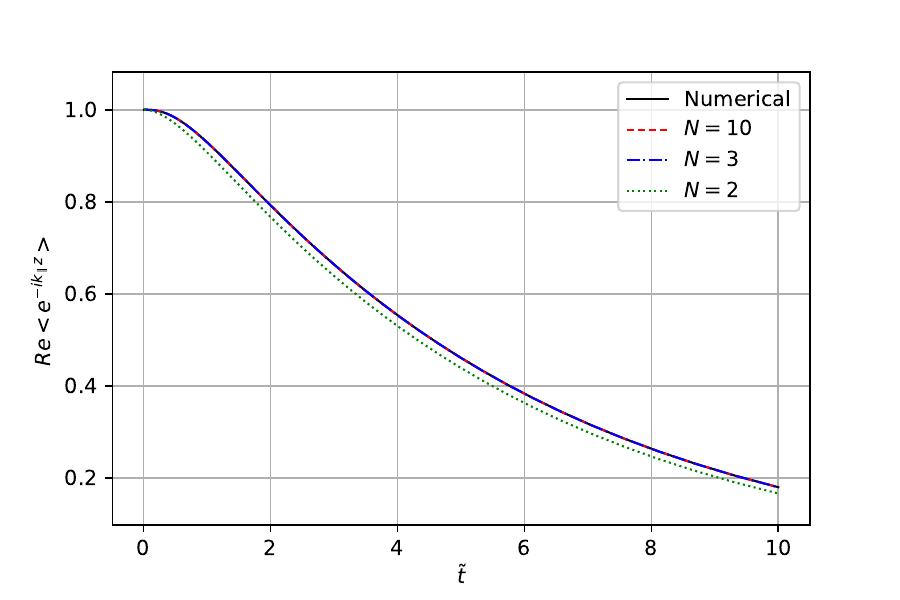}
\caption{Caption is as in Figure \ref{modified_char_5_k_01} except that we have used $\tilde k =1$.}
\label{modified_char_5_k_1}
\end{figure}

\begin{figure}[H]
\centering
\includegraphics[scale=0.5]{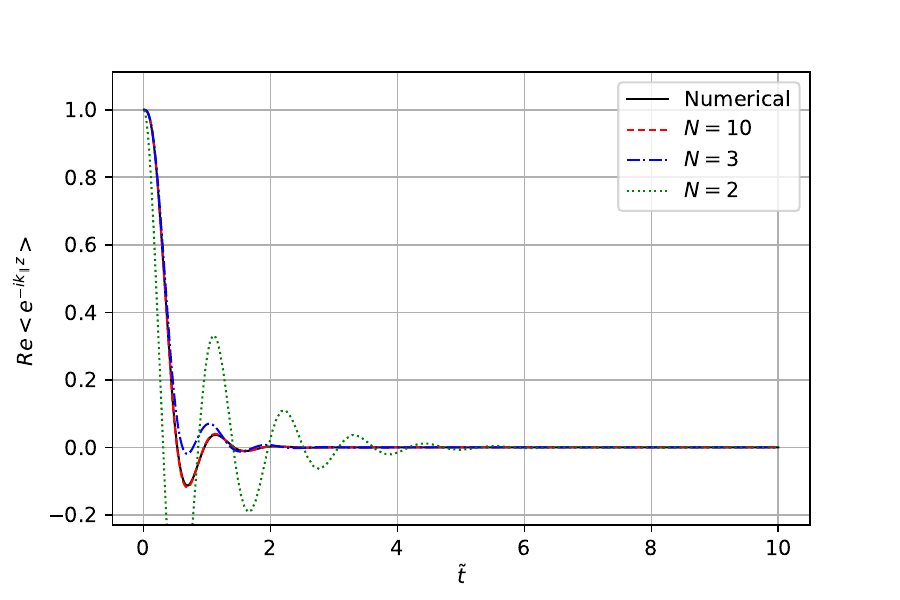}
\caption{Caption is as in Figure \ref{modified_char_5_k_01} except that we have used $\tilde k =10$.}
\label{modified_char_5_k_10}
\end{figure}

\begin{figure}[H]
\centering
\includegraphics[scale=0.5]{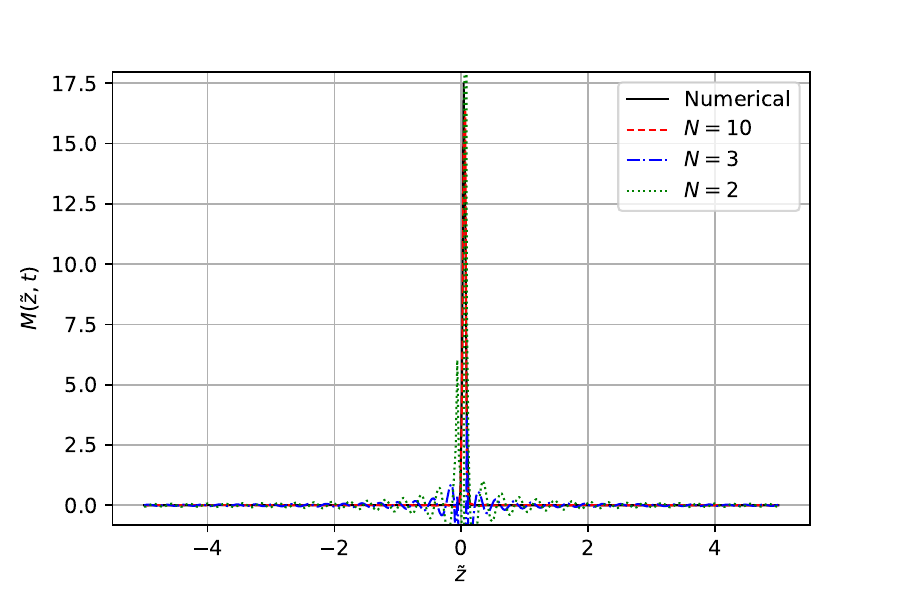}
\caption{This plot compares the $\mu$-integrated solution of the subspace method for various dimensions and the numerical solution at the time $\tilde t=1/10$
for an initial value of $\mu_0=1/2$. We have set $\xi  = 1$.}
\label{modified_mu_integrated_1_01}
\end{figure}

\begin{figure}[H]
\centering
\includegraphics[scale=0.5]{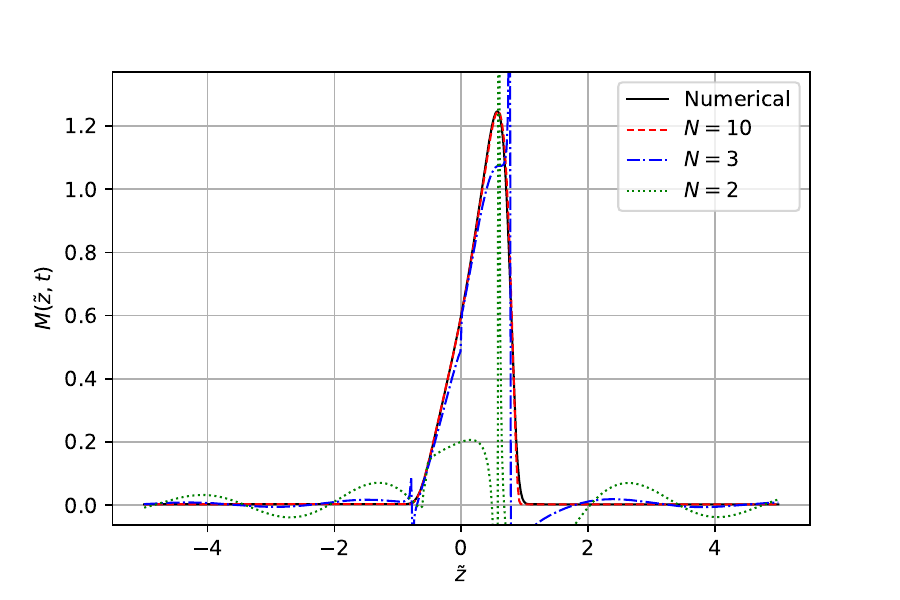}
\caption{Caption is as in Figure \ref{modified_mu_integrated_1_01} that we have used $\tilde t =1$.}
\label{modified_mu_integrated_1_1}
\end{figure}

\begin{figure}[H]
\centering
\includegraphics[scale=0.5]{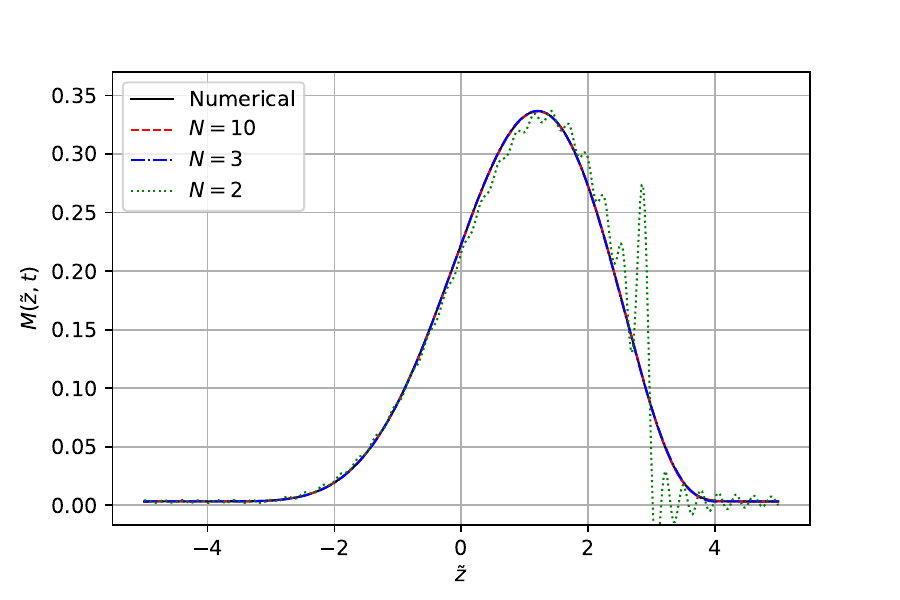}
\caption{Caption is as in Figure \ref{modified_mu_integrated_1_01} that we have used $\tilde t =5$.}
\label{modified_mu_integrated_1_5}
\end{figure}

\begin{figure}[H]
\centering
\includegraphics[scale=0.5]{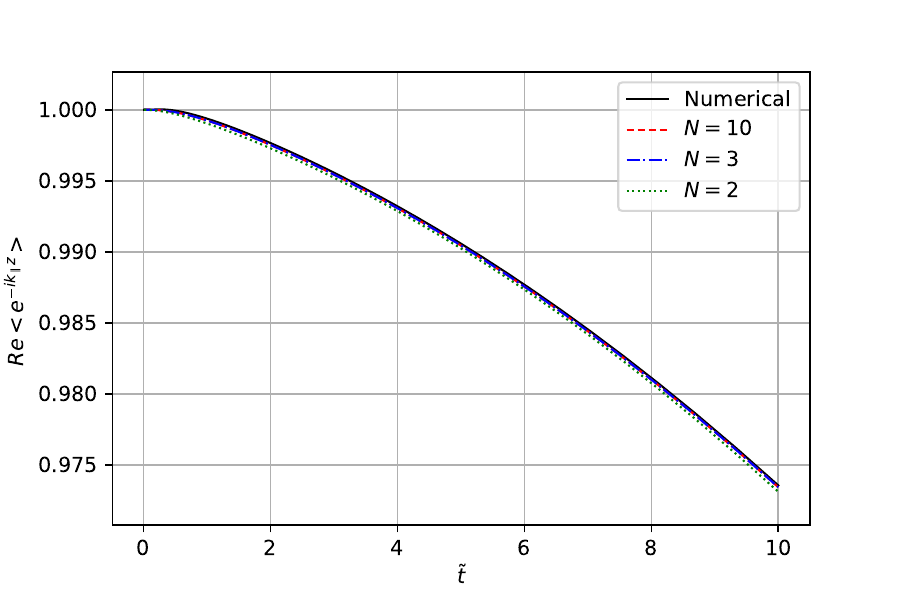}
\caption{Comparison of the $\mu$-averaged characteristic function between the subspace method for various dimensions and the numerical solution.
We have used an initial value of $\mu_0=0$ and set $\xi =1$. Furthermore, we kept the dimensionless wave number $\tilde k= vk/D=1/10$ constant and varied time.}
\label{modified_char_1_k_01part2}
\end{figure}

\begin{figure}[H]
\centering
\includegraphics[scale=0.5]{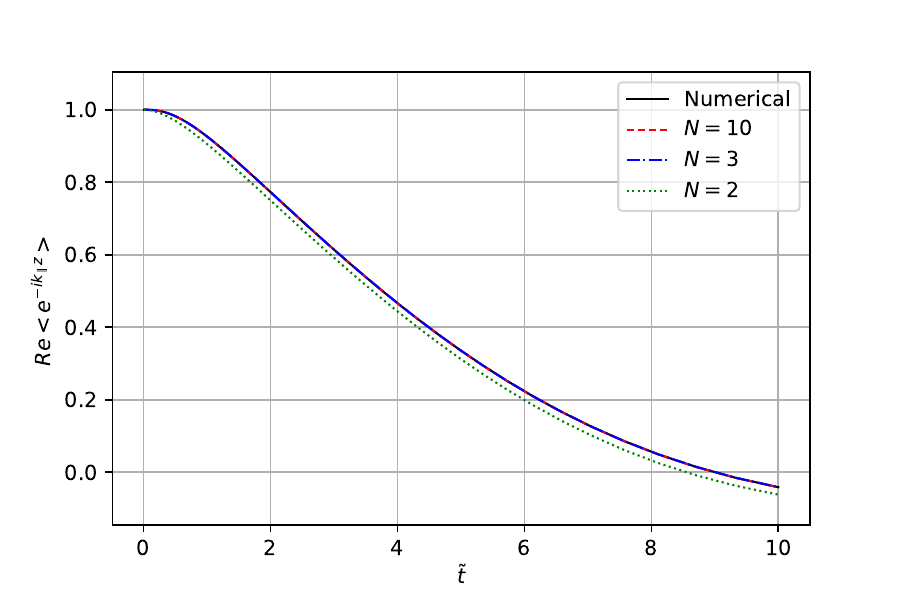}
\caption{Caption is as in Figure \ref{modified_char_1_k_01part2} but we have set $\tilde k =1$.}
\label{modified_char_1_k_1}
\end{figure}

\begin{figure}[H]
\centering
\includegraphics[scale=0.5]{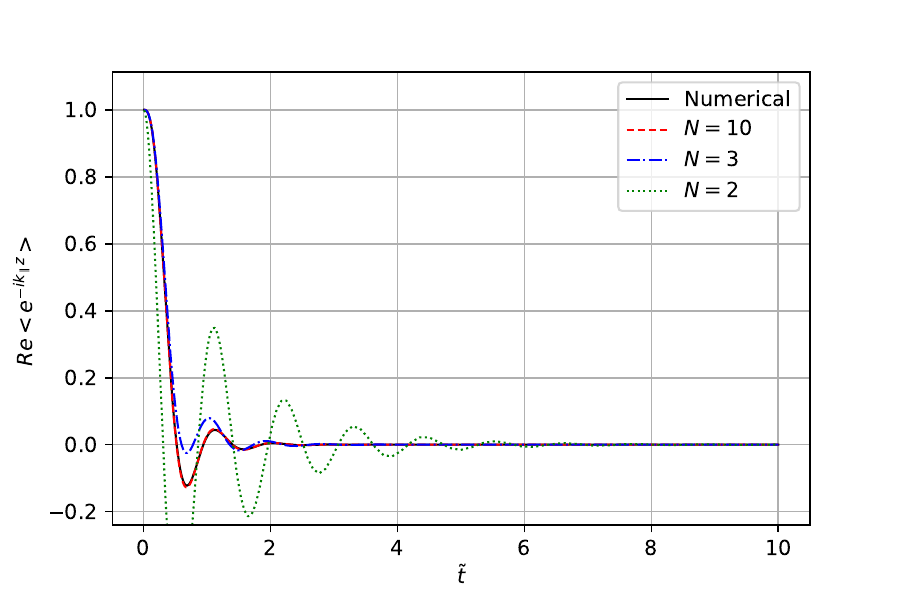}
\caption{Caption is as in Figure \ref{modified_char_1_k_01part2} but we have set $\tilde k =10$.}
\label{modified_char_1_k_10}
\end{figure}

\begin{figure}[H]
\centering
\includegraphics[scale=0.5]{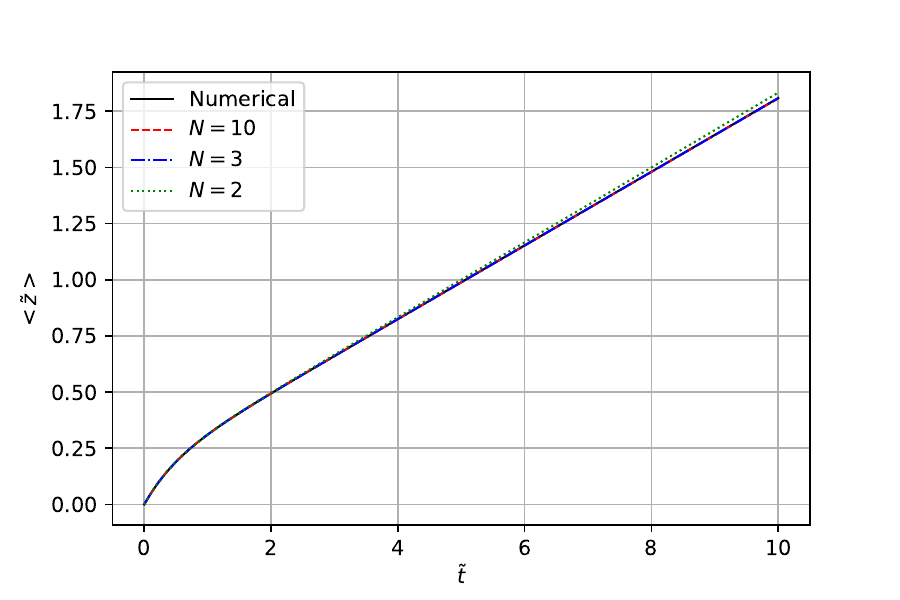}
\caption{Comparison of the mean position $\tilde z= Dz/v$ as a function of time $\tilde t$ for the $N$-dimensional subspace method and the numerical solution.
Here, we have set $\mu_0=1/2$ and $\xi =1$.}
\label{modified_Expectzxi1}
\end{figure}

\begin{figure}[H]
\centering
\includegraphics[scale=0.5]{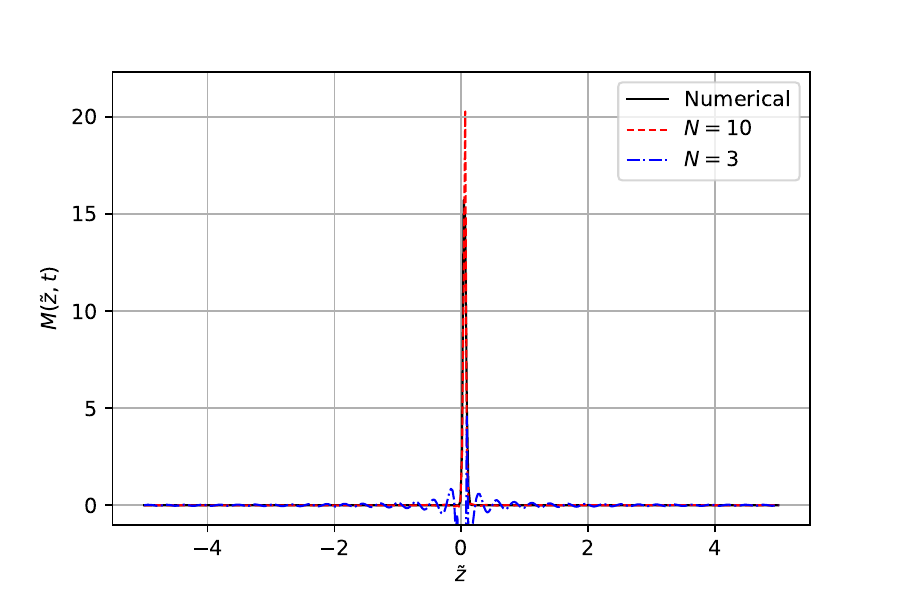}
\caption{The graph compares the $\mu$-integrated solution of the subspace method for various dimensions and the numerical solution at the time $\tilde t=1/10$
for an initial value of $\mu_0=1/2$. We have set $\xi  = 5$.}
\label{modified_mu_integrated_02_01}
\end{figure}

\begin{figure}[H]
\centering
\includegraphics[scale=0.5]{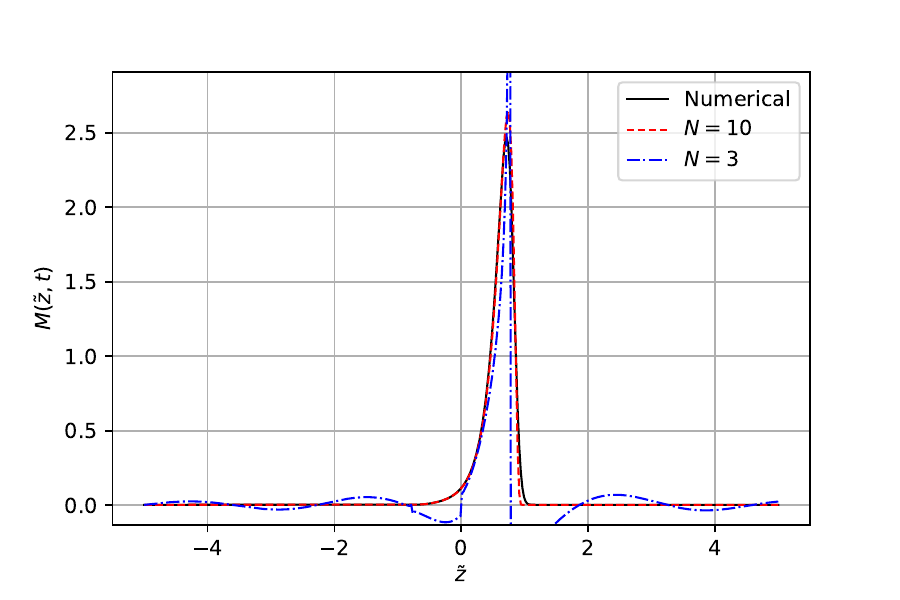}
\caption{Caption is as in Figure \ref{modified_mu_integrated_02_01} but we have used $\tilde t =1$.}
\label{modified_mu_integrated_02_1}
\end{figure}

\begin{figure}[H]
\centering
\includegraphics[scale=0.5]{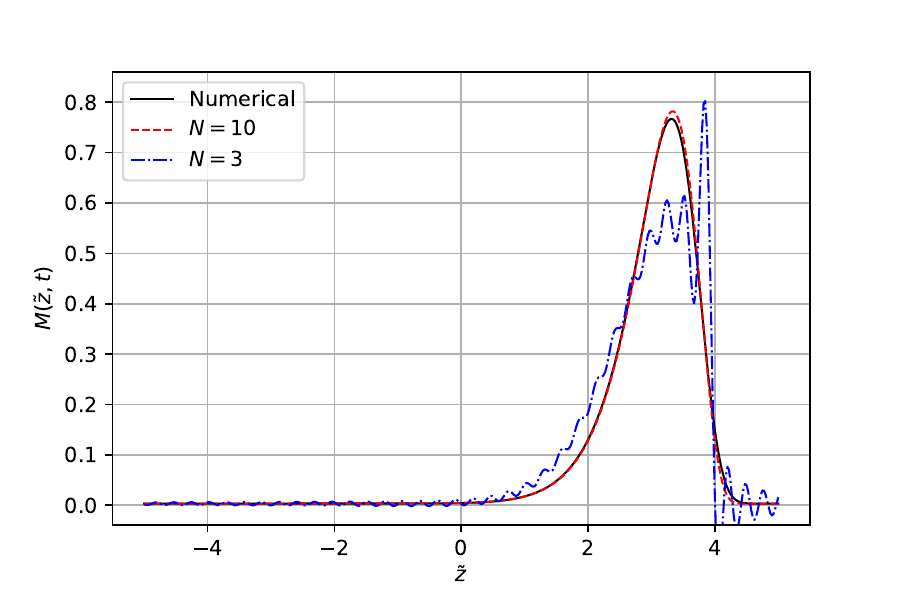}
\caption{Caption is as in Figure \ref{modified_mu_integrated_02_01} but we have used $\tilde t =5$.}
\label{modified_mu_integrated_02_5}
\end{figure}

\begin{figure}[H]
\centering
\includegraphics[scale=0.5]{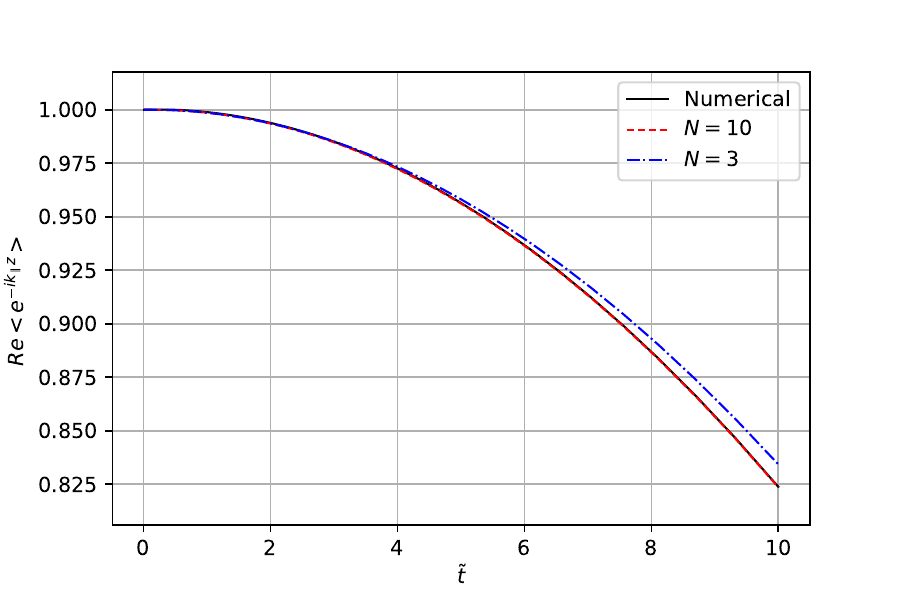}
\caption{Comparison of the $\mu$-averaged characteristic function between the subspace method for various dimensions and the numerical solution.
We have used an initial pitch-angle cosine of $\mu_0=0$ and for the focusing parameter we used $\xi =5$. We have kept the dimensionless wave number constant
at $\tilde k= vk/D=1/10$ and varied time.}
\label{modified_char_02_k_01}
\end{figure}

\begin{figure}[H]
\centering
\includegraphics[scale=0.5]{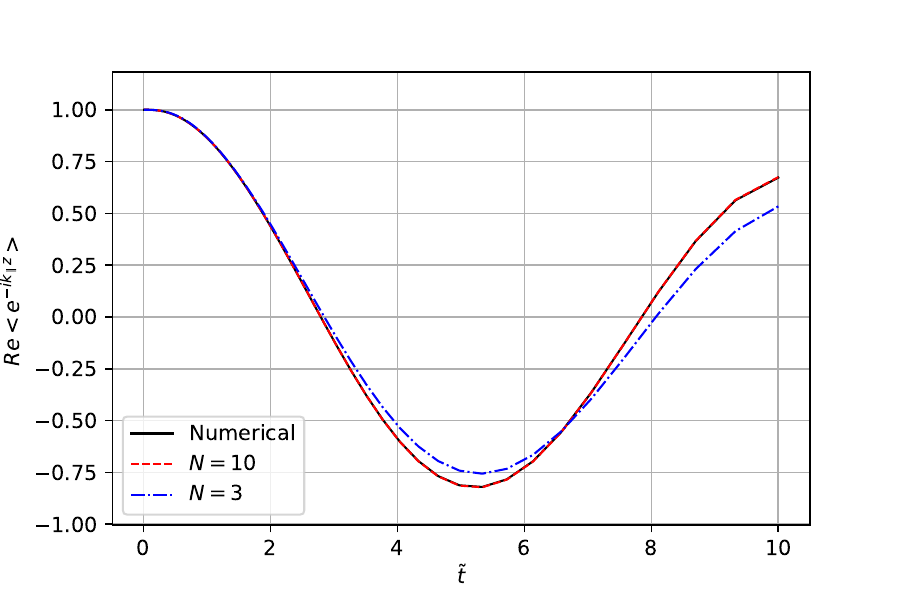}
\caption{Caption is as in Figure \ref{modified_char_02_k_01} but we have used $\tilde k =1$.}
\label{modified_char_02_k_1}
\end{figure}

\begin{figure}[H]
\centering
\includegraphics[scale=0.5]{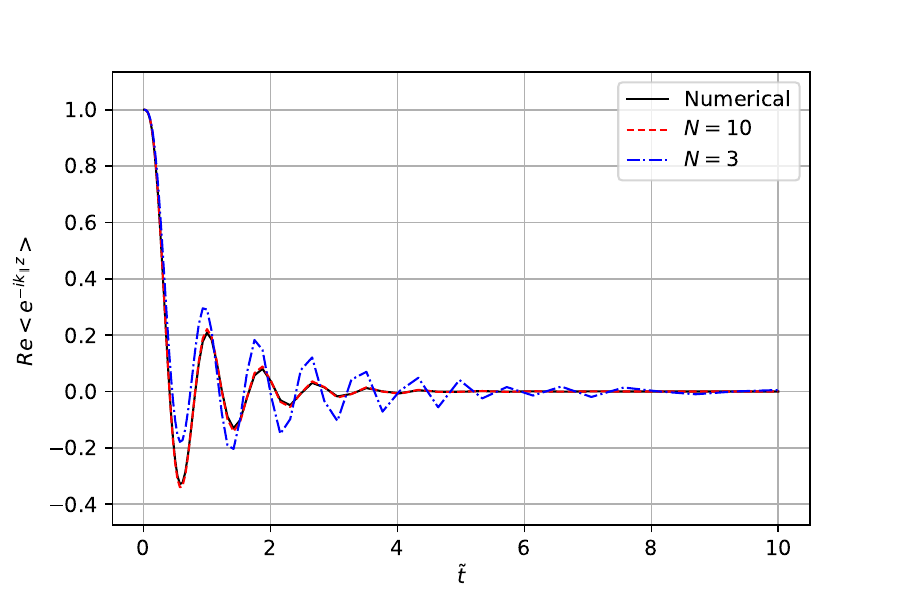}
\caption{Caption is as in Figure \ref{modified_char_02_k_01} but we have used $\tilde k =10$.}
\label{modified_char_02_k_10}
\end{figure}

\begin{figure}[H]
\centering
\includegraphics[scale=0.5]{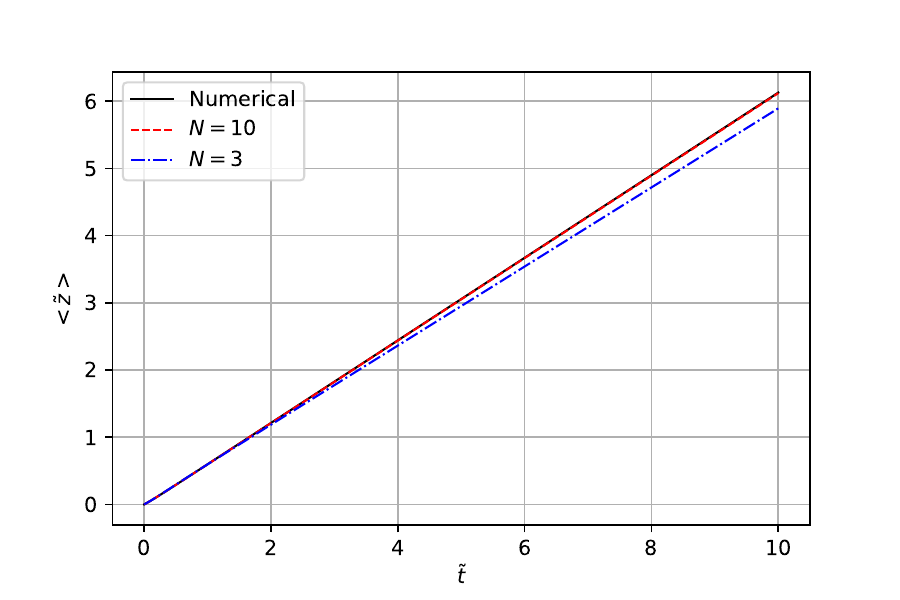}
\caption{Comparison of the mean position $\tilde z= Dz/v$ as a function of time $\tilde t$ for the $N$-dimensional subspace method and the numerical solution.
Here, we have used $\mu_0=1/2$ and $\xi =5$.}
\label{modified_Expectzxi5}
\end{figure}

The obtained results are visualized via Figs. \ref{modified_mu_integrated_5_01}-\ref{modified_Expectzxi5}. As in Paper I we decided to work with the dimensionless
quantities
\be
\tilde{t} = D t \quad\textnormal{and}\quad \tilde{z} = \frac{D z}{v}
\ee
transforming Eq. (\ref{MFP2}) into
\be
\frac{\d f}{\d \tilde{t}} +\mu \frac{\d f}{\d \tilde{z}} = \frac\d{\d\mu} \left[ \left( 1 - \mu^2 \right) \frac{\d f}{\d\mu} \right]
- \frac{1}{2} \xi \frac{\partial}{\partial \mu} \left[(1-\mu^2)\tilde{f}\right]
\label{FourierFPeqTrans}
\ee
with the dimensionless parameter
\be
\xi = \frac{v}{D L}.
\ee
Note, sometimes the \textit{focusing parameter} is defined slightly different as $\xi = v /(2 D L)$ (see, e.g., Shalchi \& Klippenstein (2025)).

In the shown figures we have compared the pure numerical solutions with solutions obtained by employing the subspace method for the cases $N=2$, $3$, and $10$. 
As in Paper I we conclude that the case $N=10$ is the best choice for a fast but still very accurate solution. 

We also compare the subspace method with the numerical solution in terms of various expectation values. Rather than computing a given expectation value
directly from the definition given in Eq. (\ref{defaverageoperator}), see Section 4 for analytical results which render the computations significantly quicker.

\begin{figure}[H]
\centering
\includegraphics[scale=0.5]{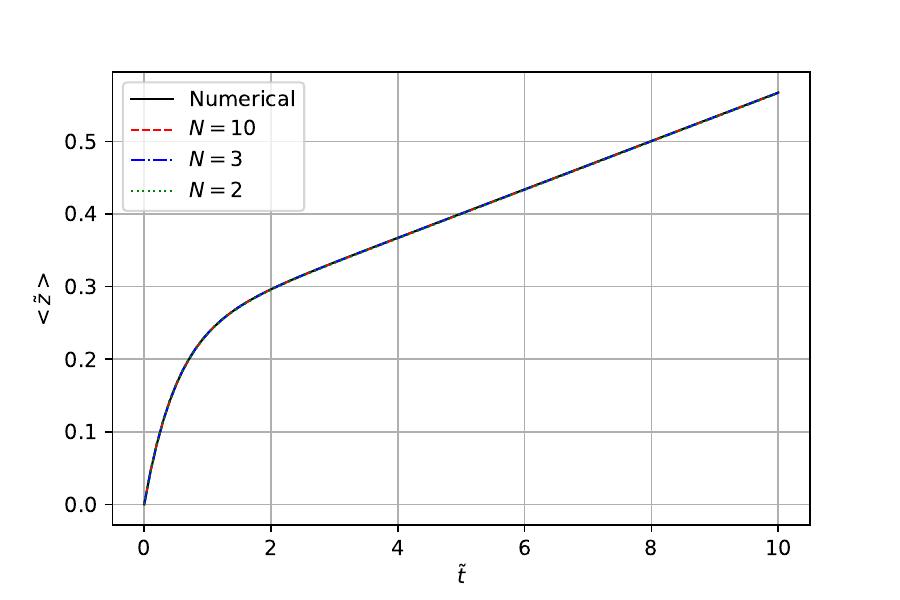}
\caption{Comparison of the mean position $\tilde z= Dz/v$ as a function of time $\tilde t$ for the $N$-dimensional subspace method and the numerical solution.
Here, we have used $\mu_0=1/2$ and $\xi =1/5$.}
\end{figure}

\subsection{Time Comparisons}

As in Paper I, we compare the time required to compute the solution using the different methods we have discussed. Tables \ref{modified_time02}-\ref{modified_time5}
show that in terms of the runtime across the different methods, the Fokker-Planck equation is very similar compared to the modified Fokker-Planck equation.

\begin{table*}[h]
\centering
\begin{tabular}
{ |p{3cm}|p{3cm}|p{3cm}|p{3cm}|p{3cm}| }
 \hline
$Dt$ &2D &3D & 10D & Numerical\\
\hline
 $0.1$&$1127.9$&$1777.1$&$6807.8$&$6117.3$\\
\hline
$1$&$34.9$&$55.2$&$215.6$&$1812.0$\\
\hline
$5$&$3.1$&$4.8$&$18.5$&$778.7$ \\
\hline
\end{tabular}
\caption{For different values of $Dt$,  we compare the time in seconds to compute the solution using the subspace approximation for various dimensions with the numerical solution. Here,  we have $\xi=1/5$.}
\label{modified_time02}
\end{table*}

\begin{table*}[h]
\centering
\begin{tabular}
{ |p{3cm}|p{3cm}|p{3cm}|p{3cm}|p{3cm}| }
 \hline
$Dt$ &2D &3D & 10D & Numerical\\
\hline
 $0.1$ &$1103.4$&$1758.5$&$6771.8$&$6077.8$\\
\hline
$1$&$34.8$&$54.9$&$214.1$&$1800.4$\\
\hline
$5$&$3.0$&$4.7$&$18.4$&$773.7$\\
\hline
\end{tabular}
\caption{Caption is as in Table \ref{modified_time02}, except that here we have used $\xi=1$.}
\end{table*}

\begin{table*}[h]
\centering
\begin{tabular}
{ |p{3cm}|p{3cm}|p{3cm}|p{3cm}| }
 \hline
$Dt$ &3D & 10D & Numerical\\
\hline
$0.1$&$1730.2$&$6759.3$&$6066.9$\\
\hline
$1$&$56.3$&$216.3$&$1810.3$\\
\hline
$5$&$4.9$&$19.2$&$778.0$\\
\hline
\end{tabular}
\caption{Caption is as in Table \ref{modified_time02}, except that here we have used $\xi=5$.}
\label{modified_time5}
\end{table*}

\subsection{Limitations of the $N$-Dimensional Subspace Approximation}\label{flaw}
It is important to point out that the $N$-dimensional subspace approximation has convergence issues which are dependent on the value of $\xi$. To understand this,
we evidently need the integrand in Eq.  (\ref{FourierTransform}) to converge to zero for large wave numbers. By Eq. (\ref{tildecn_exact}), we can see that convergence
of the solution in Fourier space requires the eigenvalues of the matrix $\textbf{M}$ to have a negative real component. However, depending on the values of $\xi$ and $N$,
some of the eigenvalues might have a positive real component.  

We explain these limitations in more detail in Paper I when dealing with the standard Fokker-Planck equation. Moreover, the limitations present when dealing with the
modified Fokker-Planck equation are identical compared to those when working with the standard equation. The main takeaway is that one must be cautious concerning
these issues when dealing with larger values of $\xi$. However, this is highly unlikely to pose a practical issue in the context of the solar wind, as the analysis
by \cite{masung78} revealed that $\xi$ never exceeded a value of $2$ throughout $30$ events.

\section{Summary and Conclusion}
The study of particle transport represents a fundamental issue in both space physics and astrophysics.  In addition,  the outcomes of particle transport has a
variety of applications. Examples consist of increased understanding of laboratory plasma, diffusive shock acceleration, and solar modulation (\cite{zank00a},
\cite{li03}, \cite{zank04}, \cite{li05}, \cite{zank06}, \cite{doshal10}, \cite{lietal12}, \cite{ferretal14}, \cite{hu17}, \cite{shenqin18}, \cite{EngWol20},
\cite{moloto20}, \cite{EngMol21}, \cite{Shen21}, \cite{Ngobeni22}, and \cite{schwander24}).

The motion of energetic particles through turbulent magnetic fields can be described by solving transport equations such as the focused Fokker-Planck equation.
This equation is generally solved using numerical methods which result in lengthy runtimes. Subspace approximation methods have been proposed in the past, (see \cite{zank00b},
\cite{shalchi24}, and Paper I), which are considerably quicker than numerical methods. It was the goal of this paper to investigate this method when dealing with the focused
Fokker-Planck equation in conservative form. We began with deriving an approximate analytical solution using the $2$-dimensional subspace approximation. This analytical
solution can be used to obtain several approximate expectation results which have a pleasant form. The second main result here is that the $10$-dimensional subspace
approximation can indeed be used to solve Eq. (\ref{MFP2}) quickly yet accurately. The only restriction with this method is that there is a limit on how large $\xi=v/(DL)$
can be in order for the $10$-dimensions to be reliable, as discussed in more detail in Paper I. Note, in the formal limit $N \rightarrow \infty$ the subspace
method provides the exact solution of the considered transport equation. Therefore, one can start with a certain value of $N$ (such as $N=10$) and then increase this
value step-by-step until the solution does not change anymore. The so found solution is then the exact solution.

It is important to note that pitch-angle scattering is determined by the pitch-angle scattering coefficient $D_{\mu\mu}$. Throughout this work, as well as in Paper I, we
have used what is referred to as the isotropic form which is a valid in the range of intermediate to strong turbulence. Additional forms based on second-order approximations
have been derived in \cite{Shalchietal09}, which are valid for weaker turbulence. An assumption that is commonly employed when deriving different forms of $D_{\mu\mu}$ is
that the spatial variation of the mean magnetic field has no influence. However, more recent work by \cite{Tautzetal14} and \cite{florinski24} have shown that this
approximation is not fully justified. Consequently, solving the Fokker-Planck equation quickly using a method such as the subspace method with a more accurate form
of $D_{\mu\mu}$ may be of interest in future work. 

\begin{acknowledgments}
{\it Support by the Natural Sciences and Engineering Research Council (NSERC) of Canada is acknowledged.}
\end{acknowledgments}


%
%
%
{}

\end{document}